\documentclass[11pt,a4paper]{article}
\pdfoutput=1
\usepackage{graphicx}
\usepackage{color}
\usepackage{multirow}
\usepackage{amsmath}
\usepackage{textcomp}
\usepackage{amssymb}
\usepackage{slashed}
\usepackage{epsfig}
\usepackage{hyperref}
\usepackage{geometry}
\newcommand{\T}{\rule{0pt}{3.0ex}}
\newcommand{\B}{\rule[-2.2ex]{0pt}{0pt}} 
\newcommand{\eps}{\epsilon}
\newcommand{\gpZ}{g g \to \gamma Z g}
\newcommand{\gpZpp}{p p  \to \gamma Z j +X}
\newcommand{\ppg}{g g  \to \gamma \gamma g}
\newcommand{\ppZ}{g g  \to \gamma \gamma Z}

\begin{document}

\title{\bf{Production of $\gamma Z g$ and associated processes via gluon fusion at hadron colliders \\}}
\author{{Pankaj Agrawal\footnote{email:agrawal@iopb.res.in}}~~and
       {Ambresh Shivaji\footnote{email:ambresh@iopb.res.in}} \\
{Institute of Physics, Saink School Post, Bhubaneswar 751 005, India} }

\date{}

\maketitle{}

\begin{center}
 {\bf Abstract}
\end{center}
 {We present results for the process $gg \to \gamma Z g$ within 
 the standard model at hadron colliders. Due to a large gluon flux at the LHC and the 
 proposed HE-LHC, this process can have significant cross section at these machines. 
 Several thousand such events have already been produced at the LHC according to the 
 standard model predictions. This process can also be a background to new physics 
 signals. We discuss some of the important features of this process related to the 
 structure of its amplitude. We also compute the total cross section at the 
 hadron colliders and give some important kinematic distributions. 
 A comparison of our results is made with the corresponding 
 NLO calculation using the MCFM program. We briefly discuss the production of
 $\gamma \gamma g $ and $\gamma \gamma Z$ via gluon fusion. The amplitudes of 
 these processes are closely related to the $gg \to\gamma Z g$ amplitude.}

\section{Introduction}

  The quest for new physics signals is continuing at the Large Hadron Collider (LHC). The collider
  is presently running at 8 TeV centre-of-mass energy and more than 10 fb$^{-1}$ of data has been collected so far. 
  The standard model (SM) is in excellent agreement with the collected data \cite{talks}. The recent discovery 
  of a fundamental boson in the mass range of 125-126 GeV might confirm the existence of the long sought Higgs
  boson of the SM \cite{LHCexp}. With the LHC running in the multi-TeV range, the gluon-gluon scattering processes 
  that at the leading order proceed via quark loop diagrams can be important. Observation of such processes
  can be a significant test of the  SM predictions. Many of such processes have been computed in the past. 
  Processes with two or more electroweak vector bosons in the final state
  have been studied extensively in the literature \cite{vanderBij:1988fb,Glover:1988rg,Glover:1988fe,Binoth:2005ua,Binoth:2006mf,
  Campbell:2011bn,Campbell:2011cu,deFlorian:1999tp,Agrawal:1998ch,Agrawal:2012sq,Melia:2012zg,Kauer:2012hd,Agrawal:2012df}. 
   In this paper, we study the $\gamma Z g$ production via gluon fusion in detail. The process $\gpZ$ contributes
   to the cross section of the process $pp \to \gamma Zj + X$ at the next-to-next-to-leading order (NNLO) in 
   $\alpha_s$. The amplitude of this process, which occurs at the one-loop level,
   is finite and gauge invariant. Partial 
   results for this process  were reported in \cite{Agrawal:2012sq,Agrawal:2012df}. We found that its 
   contribution at the LHC, compared to the corresponding tree-level contribution, was about 4-5 $\%$
   for a given set of kinematic cuts applied on the final state particles.
   This percentage would increase at higher centre-of mass energy machines such as HE-LHC. 
   In this detailed study, we check the decoupling of the top quark at the amplitude-squared 
   and at the total cross section level. We quantify the contribution of vector and axial-vector 
   parts of the $\gamma Z g$ amplitude towards the total cross section. We also study the 
   scale dependence and the effect of choosing a specific parton distribution set. In addition,
   we consider the processes $\ppg$ and $\ppZ$ and briefly discuss their results at the LHC.
   The production of $\gamma \gamma g$ via gluon fusion has already been discussed in \cite{deFlorian:1999tp,Agrawal:1998ch}.
   We update its cross section at the LHC including the effect of top quark in the loop.
   The next-to-leading order (NLO) QCD corrections to the hadronic processes, $pp \to \gamma Zj + X$ and 
   $pp \to \gamma \gamma Z + X$ are presented in \cite{Campbell:2012ft,Bozzi:2011en}.
   In the next section, we discuss the structure of the full amplitude. 
   The method of calculation and numerical checks are presented in the section 3. 
   In the section 4, we present our numerical results. We conclude in the last section.

\begin{figure}[h!]
\begin{center}
 \includegraphics[width=0.7\textwidth]{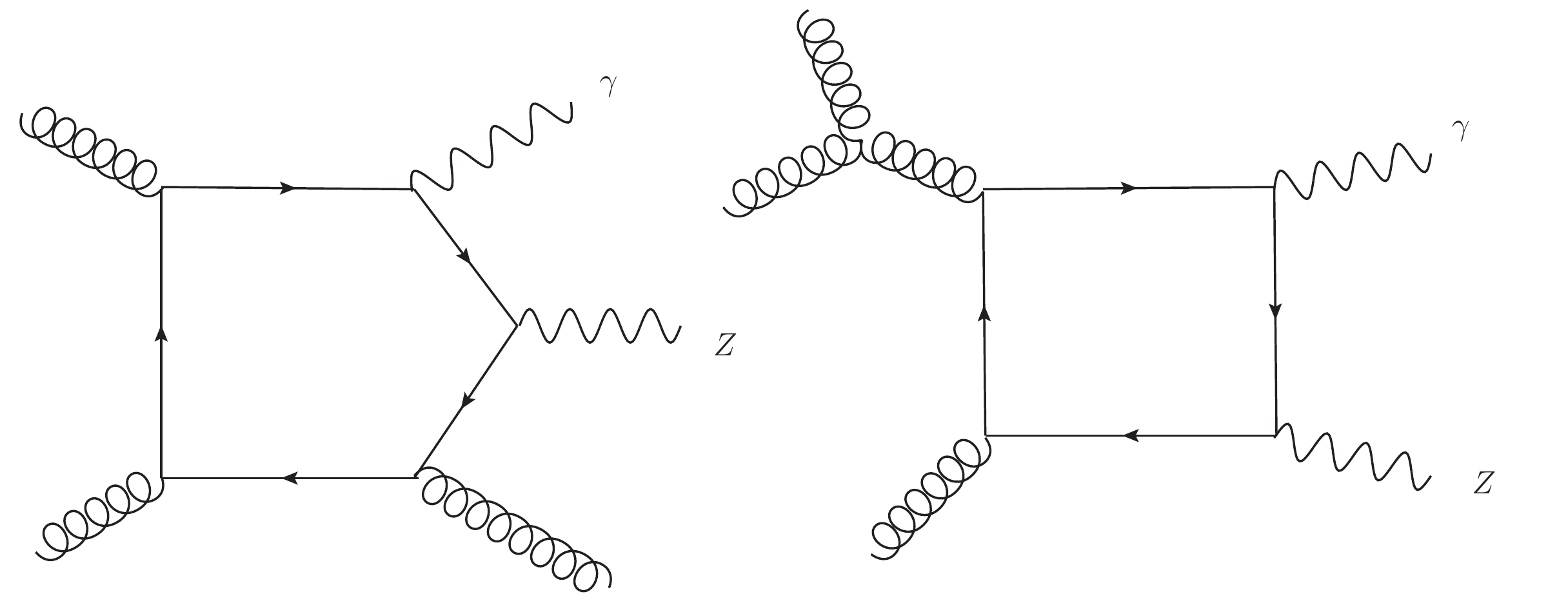}
\caption{The prototype diagrams for the processes $gg \to g \gamma Z $.}
\label{fig:gpZ}
\end{center}
\end{figure}

\section{The Amplitude}

   This one-loop process $\gpZ$ receives contribution from two main classes of quark-loop diagrams
   \textemdash  pentagon type and box type, as shown in Fig. (\ref{fig:gpZ}). The box class of diagrams are due 
   to the triple gluon vertices and can be further divided into three subclasses. This subclassification
   has its  own physical importance as we will see later. Other diagrams can be obtained by a suitable 
   permutation of external legs. For each quark flavor, there are 24 pentagon-type and $3\times6$ (= 18) 
   box-type diagrams. However, only half of the 42 diagrams are independent. Using Furry's theorem, we 
   can see that axial-vector contributions of the box diagrams add up to zero. Therefore, the box diagrams 
   give only vector contribution. The pentagon diagrams give both vector and axial-vector contributions. 
   We work with all the six quark flavors. Except the top quark, we treat all the other quarks as massless. 
  The amplitude of our process has a very nice structure. This is explained in detail in {\cite{Agrawal:2012sq}.
  The full amplitude, after including contributions from all the six quark flavors,  has following form 
\begin{eqnarray}
{\cal M}^{abc}(\gpZ) & = & i \frac{f^{abc}}{2} {\cal M}_V + \frac{d^{abc}}{2} {\cal M}_A, \\
{\cal M}_V = {\cal A}_P^V - {\cal A}_B^V & = & - \frac{e^2g_s^3}{{\rm sin}{\theta_w} {\rm cos}{\theta_w}} 
             \Bigg[\left(\frac{7}{12} - \frac{11}{9} {\rm sin}^2{\theta_w}\right) {\cal M}_V^{(0)} \nonumber \\                              
 & &       + \left(\frac{1}{6} - \frac{4}{9} {\rm sin}^2{\theta_w}\right) {\cal M}_V^{(t)}\Bigg],\\ 
{\cal M}_A = {\cal A}_P^A & = & \frac{e^2g_s^3}{{\rm sin}{\theta_w} {\rm cos}{\theta_w}} 
\left[\frac{7}{12} {\cal M}_A^{(0)} + \frac{1}{6} {\cal M}_A^{(t)}\right].
\end{eqnarray}
Here ${\cal M}_{V,A}$ are color-ordered amplitudes for the vector and axial-vector parts of the amplitude.
${\cal A}_B^V\;{\rm and}\; {\cal A}_P^{V,A}$ include full box and pentagon contributions for all the six quark flavors.
${\cal M}_{V,A}^{(0)}$ and ${\cal M}_{V,A}^{(t)}$ represent a light quark and the top quark contributions respectively. 
It is clear from 
the structure of the amplitude that the vector and axial-vector contributions are separately gauge invariant.
Therefore, the cross section of the process is an incoherent sum of the vector and axial-vector contributions.

The process $\ppg$ proceeds via pentagon and box types of quark-loop diagrams and the amplitude 
is purely vector type. On the other hand, the process $\ppZ$ proceeds via pentagon type quark-loop diagrams only and its 
amplitude is purely axial-vector type. The structure of full amplitudes for these processes are 
\begin{eqnarray}
{\cal M}^{abc}(\ppg) & = & i \; \frac{f^{abc}}{2} \; {\cal M}_V,  \\
{\cal M}_V = {\cal A}_P^V - {\cal A}_B^V & = & - e^2g_s^3 \; \left[\left(\frac{11}{9}\right) {\cal M}_V^{(0)}
                                                 + \left(\frac{4}{9} \right) {\cal M}_V^{(t)}\right].
\end{eqnarray}
\begin{eqnarray}
{\cal M}^{ab}(\ppZ) & =  & \frac{\delta^{ab}}{2} \; {\cal M}_A, \\
{\cal M}_A = {\cal A}_P^A & = & \frac{e^3g_s^2}{{\rm sin}{\theta_w} {\rm cos}{\theta_w}} \left[\left(\frac{5}{36}\right) {\cal M}_A^{(0)} 
                                                                                       + \left(\frac{1}{9}\right) {\cal M}_A^{(t)}\right].
\end{eqnarray}
 Our goal is now to compute ${\cal M}_V$ and ${\cal M}_A$ for the light quarks and the top quark loops. The
  procedure is outlined in the next section.

\section{Calculation and Numerical Checks}

 It would be sufficient to focus on the calculation of $\gpZ$ amplitude; the other two processes 
 can be considered as special cases of this process with appropriate changes in color and coupling factors.
 Our calculation is based on the traditional Feynman diagram method.
 We write down the prototype amplitudes for each class of diagrams
 using the SM Feynman rules. The amplitudes of all other diagrams are generated by appropriately
  permuting the external momenta and polarizations in our code. Only one prototype pentagon amplitude 
  is sufficient if we keep $Z$ boson external leg fixed in all permutations.\footnote{This is particularly
  economical when considering the axial-vector contributions for a massive quark case. }
 The quark-loop traces are calculated 
 using FORM \cite{Vermaseren:2000nd} in $n$ dimensions for the vector contributions. For the 
 axial-vector contribution, the quark-loop trace is calculated in 4 dimensions. 
 Interestingly, it turns out that the amplitude for the top quark in the loop and
 a massless quark in the loop can be written in such a way that they have same
 analytical form in terms of tensor integrals. The dependence on the top quark mass
 enters only through the reduction of tensor integrals. This is of considerable importance because
 it reduces the length and evaluation time of the amplitude with the top quark in the loop.  
In the case of pentagon-type diagrams, the most complicated integral
is rank-5 tensor integral ($E^{\mu \nu \rho \sigma \delta}$); while
for the box-type diagrams, rank-4 tensor integral ($D^{\mu \nu \rho \sigma}$) is
the most complicated one. 
Five point tensor and scalar integrals are written in terms of box tensor and 
scalar integrals using 4-dimensional Schouten identity.
The box tensor integrals are reduced into the standard scalar integrals -- $A_0$, $B_0$, $C_0$ and $D_0$
 following the reduction scheme
developed by Oldenborgh and Vermaseren \cite{vanOldenborgh:1989wn}. These reduction procedures were implemented
using Mathematica. We then obtain FORTRAN routines that compute various tensor integrals in terms of lower-rank 
tensor integrals and scalar integrals.
The required scalar integrals  are calculated using OneLOop library \cite{vanHameren:2010cp}.
Because of a very large and complicated expression of the amplitude, we calculate the amplitude numerically before 
squaring it. This requires numerical evaluation of the polarization vectors of gauge bosons.
We choose the real basis, instead of the helicity basis for the polarization vectors to calculate the amplitude.
This is to reduce the size of the compiled program and the time taken in running the code. The three-body
phase space is generated using RAMBO \cite{Kleiss:1985gy}.

The amplitude for $\gpZ$ is a one-loop amplitude at the LO itself. Although individual diagrams 
may be ultraviolet (UV) and/or infrared (IR) divergent, the full amplitude should be 
both UV as well as IR finite. The IR divergence is relevant to the massless quark cases only.
All these singularities are encoded in various scalar integrals. To make UV and IR 
finiteness checks on our amplitude we have derived all the required scalar integrals 
analytically following 't Hooft and Veltman \cite{'tHooft:1978xw}. We regulate  UV 
divergences of the scalar integrals using the  dimensional regularization and infrared 
singularities by giving a small mass to the quark, the mass regularization. Following 
are the details of various checks made on our amplitude.

\begin{enumerate}
\item {\it UV Finiteness:}
The tadpole and bubble scalar integrals ($A_0$ and $B_0$) are sources of 
UV singularity in any one-loop amplitude. For the case of massless internal lines,
$A_0$s don't appear in the tensor reduction.
For both the massive and massless quark contributions, we
have verified that the amplitude is UV finite. 
 The amplitude of
each pentagon diagram has only UV finite tensor integrals. Therefore, 
each pentagon diagram
is UV finite by itself as expected from naive power counting. The box diagrams
individually are not UV finite. Therefore, the cancellation of the divergence
in the sum of the box diagrams is an important check.  We find that the 
three classes of box diagrams are separately UV finite. 
\item {\it IR Finiteness:}
  The diagrams with massless internal quarks have mass singularities.
 In the mass regularization these singularities appear as ${\rm ln}^2(m_q^2)$ and ${\rm ln}(m_q^2)$  
 terms in the amplitude. We have checked explicitly that such terms are absent from
 the amplitude. We have verified that the IR finiteness holds for each quark loop diagram 
 of the box and pentagon type \cite{Shivaji:2010aq}.
\item {\it Ward Identities:}
  Certain mathematical identities can be obtained by replacing a polarization vector by its 
  4-momentum in any of the pentagon/box amplitudes. This way a pentagon amplitude can be written 
  as a difference of two (reduced) box amplitudes and also a box amplitude can be written 
  as a difference of two (reduced) triangle amplitudes. In the appendix, we have written these 
  identities for a prototype pentagon and a prototype box amplitude. We have verified them numerically.
  These Ward identities are important checks on individual diagrams and can also be utilized for a
  systematic study of numerical instabilities in the tensor reduction near exceptional phase space points.
\item {\it Gauge Invariance:}
As we have seen, because of the color structure, the vector and axial-vector parts of the amplitude
do not interfere and they are separately gauge invariant.
The vector part of the amplitude has gauge invariance with respect to three gluons, the photon and the $Z$ boson. 
This has been checked by replacing the polarization
vector of any of these gauge particles by its momentum ($\varepsilon^\mu(p_i) \rightarrow p^\mu_i$) 
which makes the amplitude vanish. As one would expect the pentagon and the three classes of box
contributions are separately gauge invariant with respect to
the $\gamma$ and the $Z$ boson. For each gluon, one of the three classes of box amplitudes is
separately gauge invariant and further cancellation takes place among
pentagon and the other two box contributions. The axial-vector part of the amplitude
is separately gauge invariant with respect to the all the three gluons and the photon.
We verify that due to explicit breaking of the chiral symmetry in the presence of quark mass, the axial-vector
part of the amplitude vanishes on replacing $Z$ boson polarization by its 4-momentum only 
in the $m_q \rightarrow 0$ limit.
\item{\it Decoupling of heavy quarks: }
As a consistency check, we have also verified that the vector and axial-vector parts of the amplitude vanish
in the large quark mass limit \cite{Appelquist:1974tg}. 
This feature of the amplitude is very closely related to its
UV structure. The decoupling holds for each pentagon amplitude and also for each class of box 
amplitudes. 
\end{enumerate}
Apart from making all above checks we have confirmed that the vector part of the $\gamma Zg $ amplitude (ignoring coupling factors)
reproduces the $\gamma \gamma g $ amplitude in the $M_Z\to 0$ limit within the working precision.

\section{Numerical Results}

    Based on the procedure outlined above, we can now compute the cross section and examine
  various features of the processes.
 As we have already mentioned, we compute amplitudes numerically using real polarization vectors
 for gauge bosons. For the processes $\gpZ$ and $\ppZ$, there are total 48 polarized amplitudes 
to be computed for each phase space point; while for the $\ppg$ process, there are 32 such
amplitudes.
 Given the number of diagrams, the number of polarization combination and the length of the amplitude,
 the computation becomes very time consuming. Each phase space point evaluation takes about 1.3 seconds
 on a single machine that we use.
 We, therefore, run the code in a parallel environment using AMCI package, a PVM implementation of the VEGAS
 algorithm \cite{Veseli:1997hr,pvm}. We have used more than 30 cores to run the code in parallel environment.
 Still it takes more than 12 hours to get suitable cross section which includes both massive and 
 massless quark contributions. We use ifort compiler on Intel Xeon CPU 3.20GHz 
 machines. 

We first study the importance of the diagrams with the top quark in the loop. For the process $\gpZ$, in 
Fig. (\ref{fig:ampsq_mt}), we plot the ratio of the amplitude-squared for five and six quark flavor contributions 
as a function of the top quark mass for a given phase space point. The partonic centre-of-mass energy, 
$\sqrt{s}$ is fixed at $8 M_Z$.
The vector and axial-vector contributions are plotted 
separately. The  $m_t = 4 M_Z$ corresponds to the scale above which the top quarks in the loop can not go on-shell. 
Various slope changes shown in this plot correspond to possibilities of producing one or more
final state particles via on-shell $t\bar t$ annihilation. In Fig. (\ref{fig:sigma_mt}), we see that the contribution of the top quark 
($m_t = 175 $ GeV) to the total cross section is negligible. We also see a knee in the plot at $m_t = {M_Z \over 2}$.
This corresponds to the $Z$ boson production due to $t{\bar t}$ annihilation. The top quark decouples at around 100 GeV. We have, therefore, not
included its contribution in our results presented below. The run time of our code is also reduced by $50 \%$.  
It is not surprising that the top quark decouples at such a low value for our processes. 
This is simply
because there are four/five quark propagators in each box/pentagon diagram leading to a large power
of the top quark mass in the denominator. \\

We can divide our numerical results that are presented in this paper into two categories.
 We first discuss theoretical results, related to the structure of the amplitude,
keeping the $Z$ boson on-shell. 
Theoretical results presented below include following kinematic cuts:
$
p_T^{j} > 30 \; {\rm GeV },\; p_T^{\gamma,Z} > 20 \; {\rm GeV },\:|\eta^{\gamma,Z,j}| < 2.5,\; R(\gamma,j) > 0.6 .
$
These cuts, together with $R(\gamma,\gamma) > 0.6$, are also used to compute total cross sections for $\gamma \gamma g$ and
$\gamma \gamma Z$ cases. 
We have chosen the factorization and renormalization scales as
$\mu = \mu_F = \mu_R = E_T^Z$, the transverse energy of the $Z$ boson. Results are obtained using CTEQ6M 
parton distribution functions \cite{Nadolsky:2008zw}. 
In Fig. (\ref{fig:sigma_cme-V+A}),
we present dependence of the total cross section on collider centre-of-mass energy  to see
the effect of large gluon luminosity at higher energies. We have already seen 
that the vector and axial-vector parts of the amplitude are separately gauge invariant.
Their contributions towards the total cross section are also included in the figure.
The axial-vector contribution is only about $10 \%$ of the total cross section;
this contribution comes from the pentagon class of diagrams only. Although the
box amplitude contribution to the cross section is not separately gauge invariant with respect to the gluons, it 
is gauge invariant with respect to the $\gamma$ and the $Z$ boson. We find that more than
$70 \%$ of the total cross section is due to the box amplitude only, see Fig. (\ref{fig:sigma_cme-box}).
The scale variation of the total cross section about the central value $\mu = E_t^Z$ is shown
in Fig. (\ref{fig:sigma14Q}). On increasing the scale by a factor of 2, the cross section 
decreases by about $25 \%$; it increases by about $40 \%$ on decreasing the scale by a factor of 2. 
These large variations are expected because our calculation is effectively 
LO as far as $\mu$ dependence is concerned. 
We see that the cross section falls as we increase the scale $\mu$.
This is because an increase in the factorization scale increases the cross section due to increase
in gluon luminosity; but an increase in the renormalization scale decreases the cross section
because of the decrease in  the value of $\alpha_s(\mu)$. When we increase both the scales 
at the same time, the effect of the change in renormalization scale is stronger. It leads to an
overall decrease in the cross section with the increase in the scale $\mu$.

 Next, we come to the discussion on our phenomenological results. These results include various kinematic distributions 
related to final state particles. For phenomenological results, we work in the narrow width approximation.
We allow the $Z$ boson to decay into two leptons in the phase space. Here, by lepton we mean an electron or a muon.
In this case the kinematic cuts are,

$$ p_T^{j} > 30 \; {\rm GeV },\; p_T^{\gamma} > 15 \; {\rm GeV },\; p_T^l > 10 \; {\rm GeV },\:|\eta^{\gamma,l,j}| < 2.5, \nonumber $$
$$ R(\gamma,j),R(l,\gamma),R(l,j),R(l,l) > 0.4. \nonumber $$
For convenience we have chosen the scale $\mu = \mu_F = \mu_R = M_Z$.
In Fig. (\ref{fig:sigma_cme-gpz}), we give cross section variation in the range of 8 to 14 
TeV centre-of-mass energy using both CTEQ6l1 and CTEQ6M parton distribution functions. 
These numbers do not include the branching ratio of $Z \to l^{+} l^{-}$.  In particular, 
the cross sections, with CTEQ6l1 (CTEQ6M) parton distributions, are 65.4 (53.0) fb and 
202.4 (154.3) fb at 8 TeV and 14 TeV centre-of-mass energies respectively.
However, to observe these events, one would need to detect the decay products of the
Z boson. For the best observability, one may have to look at $Z \to l^{+} l^{-}$ decay channel; here $l$ can be
 an electron/muon. 
Including this branching ratio, one may expect more than
1000, $gg \to \gamma Z (\to l^{+} l^{-}) g$ events at the 14 TeV LHC, with 100 fb$^{-1}$ integrated luminosity.
The transverse momentum and rapidity distributions for the
final state particles are shown in Figs. (\ref{fig:ptp8}-\ref{fig:etal28}) at 8 TeV centre-of-mass energy. 
We have given normalized distributions as they remain same for different choices of parton distributions
and/or scales. These distributions are characteristically similar at different collider centre-of-mass
energies, but at higher energies contribution coming from high $p_T$/rapidity region grows while low $p_T$/rapidity
region contribution goes down. 
We note that $p_T^j$ is softer as compared to $p_T^\gamma$. It is because the cross section is dominated by the
box class of diagrams and in these diagrams, the gluon is emitted as a bremsstrahlung radiation. Due
to the same reason, {\it i.e.}, the gluon is emitted more collinearly, the rapidity distribution of
the gluon is broader as compared to that of the photon. The lepton-$p_T$ distribution peaks around $M_Z/2$.
On the other hand, the rapidity distribution of the lepton is more central compared to 
the $\eta^\gamma$ distribution.
\begin{table}
\begin{center}
\begin{tabular}{|c|c|c|c|c|c|}
\hline
$\sqrt{{\rm S}}$ (TeV) & $p^{\gamma,min}_T$ (GeV) & $\sigma^{{\rm LO}}$ (pb) & $\sigma^{{\rm NLO}}$ (pb) 
& $\sigma^{{\rm NNLO}}_{gg}$ (fb) & $\sigma^{{\rm NNLO}}_{gg}/\sigma^{{\rm NLO}}(\%)$ \T \B \\
\hline 
\multirow{2}{*}{8} & 30 & 2.202 & $3.391^{+9.48\%}_{-7.03\%}$ & 46.05 (38.25) &  1.36 (1.13) \T \\
                   & 50 & 1.144 & $1.744^{+10.3\%}_{-6.22\%}$ & 30.49 (25.61) &  1.75 (1.47) \T \B \\ 
\hline
\multirow{2}{*}{14} & 30 & 4.868 & $7.722^{+12.2\%}_{-6.68\%}$ & 158.72 (124.48) & 2.06 (1.61) \T \\
                    & 50 & 2.608 & $4.158^{+7.42\%}_{-6.41\%}$ & 109.92 (86.61)  & 2.64 (2.08) \T \B \\ 
\hline
\multirow{2}{*}{35} & 30 & 14.973 & $23.548^{+8.87\%}_{-2.72\%}$ & 854.09 (606.07) & 3.63 (2.57) \T \\
                    & 50 & 8.220  & $13.514^{+6.87\%}_{-3.60\%}$ & 607.35 (438.88) & 4.49 (3.25) \T \B \\ 
\hline
\end{tabular}
\end{center}
\caption{ Cross sections for the production of $\gpZpp$ at various collider centre-of-mass energies. We use
CTEQ6l1 PDF set at LO and CTEQ6M PDF set at NLO. The NNLO predictions are with CTEQ6l1 (CTEQ6M) parton distribution.
The factorization and renormalization scales are set to $\mu_F = \mu_R = M_Z$. The percentage changes, upon varying this
scale by a factor of two, are also quoted for the NLO results. }
\end{table}
 We have also compared results of this NNLO calculation with the LO and NLO predictions for 
$ p p \to \gamma (Z \to \nu \bar\nu) j + X$ \cite{Campbell:2012ft}. The LO and NLO results are 
obtained using parton-level next-to-leading order program MCFM \footnote{http://mcfm.fnal.gov/}.
The comparison is presented after removing the branching ratios in Table 1. 
The table includes results 
at three different centre-of-mass energies and for two values of the $p_{T}^{\gamma,min}$.
We have included the centre-of-mass energy of 35 TeV, as it is proposed for the HE-LHC collider.
The other kinematic cuts are: $p_T^{j} > 30 \; {\rm GeV },\; p_T^{miss} > 30 \; {\rm GeV },\:|\eta^{\gamma,j}| < 2.5,
R(\gamma,j) > 0.4$.
This table illustrates two facts -- 1) the fraction of
NNLO events increases with the increase in  $p_{T}^{\gamma,min}$, 2) the NNLO process becomes
more important as we increase the centre-of-mass energy. There is an increase in the NNLO fraction with an increase in $p_{T}^{\gamma,min}$ 
because, in the NLO events, 
the photon is emitted from a quark line; so a larger  $p_{T}^{\gamma,min}$ suppresses the
NLO contribution more than the NNLO contribution. 
In Fig. (\ref{fig:sigma14ptp30}) we have compared the 
normalized $p_T^\gamma$-distributions at the NLO and NNLO, leading to the same conclusion. 
The importance of the NNLO process is more
at higher centre-of-mass energy simply because of the increase in the gluon-gluon luminosity.

The variation of the total cross section with the collider centre-of-mass energy for $\gamma \gamma g$ and $\gamma \gamma Z$
production via gluon fusion are given in Figs. (\ref{fig:sigma_cme-ppg}) and (\ref{fig:sigma_cme-ppz})
respectively.  We reconfirm the importance of $\ppg$ at the LHC. Like the $\gpZ$ process, we have checked 
that the diagrams with the top quark in the loop make negligible contribution. The cross section at the
14 TeV centre-of-mass energy is of the order of 2 pb, leading to several thousand events with
even 1 fb$^{-1}$ of integrated luminosity. The amplitude for $\ppZ$ is purely axial-vector
type and the coupling factors are smaller as compared to the axial-vector part of the $\gamma Zg$ amplitude.
We find that its cross section is very small at the LHC, about 0.02 fb. Therefore, this process is unlikely
to be observed at the LHC.  \\

 Like other calculations of our types, we have also faced the issue of numerical instability in our 
 calculation for certain phase space points. 
 This is a well known issue with the reduction of one-loop tensor integrals of higher rank and higher points.
 This issue is also faced in the evaluation of scalar integrals.
 We face numerical instabilities in the evaluation of pentagon-tensor integrals. This is related to the inaccurate 
 evaluation of Gramm determinants in those phase space regions  where the linear independency of 
 external momenta (modulo 4-momentum conservation) is compromised. The inverse Gramm determinants appear in 
 the reduction of tensor integrals. Near exceptional phase space points, the Gramm determinants become very small
 and give rise to numerical problems because of loss of precision due to large cancellation. 
This problem can be handled in several ways. 
 One way is to use higher precision for tensor reduction and scalar integral calculations. 
This certainly reduces 
 the number of exceptional phase space points, but the code becomes enormously slow. Another approach 
 could be to use special expressions for tensor reduction for such phase space points \cite{Denner:2005nn}. This 
    approach is some time better.
    It is important to mention here that none of these two approaches cure the problem of numerical instability 
    completely \cite{Campanario:2011cs}. A more economic and convenient way to proceed in 
    this situation is to judiciously ignore the  contributions from such phase space points. This one
    can do because we are not doing precision calculations and exceptional phase space points are very
    special and are unlikely to give a significant contribution to the total cross section.
    We perform a gauge-invariance (GI) test on full amplitude for each phase space 
    point. In practice, we put a GI test cut of, $\delta = 10^{-6}$ on amplitude-squared. 
    We ignore all those points which fail to pass this test. With 
    this cut on $\delta$ and other kinematic cuts, the number of points ignored are below $2 \% $. 
    However, with higher $p_T$ cut and/or less stringent cut on the $\delta$, the
    number of such points can drop to a level of $0.01 \%$. We sample about 0.4-0.5 million 
    phase space points to obtain the numerical results.
    Given the volume of phase space, the number of exceptional 
    phase space points is small and it is reasonable to assume that the cross section
    is not dominated by this region of phase space. We find that our result depends on 
    this cut very weakly and remains quite stable over the range of $10^{-4}-10^{-12}$ 
    for the choice of cut. This also reflects that the exceptional phase space points are
    few and make small contribution. This can be seen in Fig. (\ref{fig:sigma-cut}).
    We have also checked our results for cross section calculation by implementing 
    a set of Ward identity tests and its sensitivity on $\delta$ like cut parameter. 
    Though the total cross section is quite stable, various distributions (specially 
    the rapidity distributions near the edges) are quite sensitive to variation in 
    $\delta$. Edges of distributions define the region where exceptional phase space 
    points may lie. Therefore, inaccuracy of the distributions at the edges of
    the phase space is not surprising. We have seen that the exceptional phase space 
    points may defy the Ward identity tests sometimes. This might be related to the
    form of these identities as shown in eqns. (8) and (9) \footnote{ The actual 
    implementation of Ward identities, say for pentagon diagrams, is in the form of, 
    $(1-\frac{{\cal M}_P}{{\cal M}_B^1-{\cal M}_B^2}) < \delta$.}.  One can make these
    $\delta$-like cuts more stringent to get more reliable distributions. One can also
    identify exceptional phase space points at  the level of Gramm determinants which 
    may be more economical. Phase space points corresponding to 
    large cancellation in Gramm determinants can be ignored without 
    putting stringent cuts on $\delta$ and again leading to more reliable distributions.
    A method to implement this criterion is discussed in \cite{Campanario:2011cs}.

\section{Conclusions} 
   In this paper, we have presented the results of the cross section 
   calculations for the processes $\gpZ,\; \ppg$ and $\ppZ$ at the hadron 
   colliders. These processes occur at the one-loop level via pentagon and 
   box diagrams. We have made a number of checks on our calculations. We 
   have verified the finiteness of the amplitude and have checked its 
   gauge invariance with respect to all the gauge particles. We have also
   checked the decoupling of the heavy quark and some of the Ward identities
   which are listed in the Appendix.
   For the $\gamma Z g$ process, the axial-vector contribution does not 
   interfere with the vector contribution and it is separately gauge 
   invariant. 

    Results are obtained using the parallel running of the code.
    We have presented the cross sections and various kinematic distributions
    mainly for the $gg \to \gamma Z g$ process. We 
    find that the top quark contribution is negligible; and that the 
    vector coupling contributions dominate over the axial vector couplings.
    The typical cross section for the process $\gpZ$ is of the order of 
    $100-200$ fb. A comparison with MCFM suggests that the contribution
    of this process to the NLO cross section of $pp \to \gamma Z j$ is about
    $2-3\%$. One may be able to enhance this percentage further
    by choosing an appropriate set of kinematic cuts.
    The amplitude for $\ppg$ is purely vector type and its typical 
    cross section is about 2 pb. Both these processes should be 
    observable at the LHC.  However, the process $\ppZ$ is
    more suppressed due to it being a purely axial-vector pentagon process
    and it is unlikely to be seen at the LHC.

 \section*{Acknowledgment}
 We would like to thank people involved in maintaining the 
 small cluster facility of forty cores at the institute computer centre. 
 A.S. would like to acknowledge fruitful discussions with Francisco Campanario
on the issue of numerical instability. He also wishes to thank John M. Campbell
 and C. Williams for providing help with the MCFM program.

\section{Appendix: Ward-Identity for Pentagon and Box Amplitudes }
We wish to write down the Ward-Identity for prototype pentagon and box diagrams shown in Fig. (\ref{fig:gpZ}). We choose replacing 
$Z$ polarization vector by its 4-momentum to setup pentagon identities. We ignore all the irrelevant factors of color and coupling etc.
and consider vector part of the amplitude. It can easily be extended for axial-vector part of the amplitude. \\
$$
{\cal M}_P(p_1,p_2,p_3,p_4;\eps_1,\eps_2=p_2,\eps_3,\eps_4,\eps_5) \equiv {\rm tr} \left( 
                          \slashed \eps_5\frac{1}{\slashed k_{1234}-m_q}
                          \slashed \eps_4\frac{1}{\slashed k_{123}-m_q}
                          \slashed \eps_3\frac{1}{\slashed k_{12}-m_q}
                          \slashed p_2\frac{1}{\slashed k_{1}-m_q}
                          \slashed \eps_1\frac{1}{\slashed k-m_q} \right) \\
$$
$$
  \nonumber \\= {\rm tr} \left( \slashed \eps_5\frac{1}{\slashed k_{1234}-m_q}
             \slashed \eps_4\frac{1}{\slashed k_{123}-m_q}
             \slashed \eps_3\frac{1}{\slashed k_{1}-m_q} 
             \slashed \eps_1\frac{1}{\slashed k} \right) -   
   {\rm tr} \left( \slashed \eps_5\frac{1}{\slashed k_{1234}-m_q}
             \slashed \eps_4\frac{1}{\slashed k_{123}-m_q}
             \slashed \eps_3\frac{1}{\slashed k_{12}-m_q} 
             \slashed \eps_1\frac{1}{\slashed k-m_q} \right) \nonumber \\ \\
$$
where we have used, $ \slashed p_2 = (\slashed k_{12}-m_q) - (\slashed k_{1}-m_q)$.
\begin{eqnarray}
\Rightarrow{\cal M}_P(p_1,p_2,p_3,p_4;\eps_1,\eps_2=p_2,\eps_3,\eps_4,\eps_5) &=&  
{\cal M}_B(p_1,p_2+p_3,p_4;\eps_1,\eps_3,\eps_4,\eps_5) \nonumber \\
&& - {\cal M}_B(p_1+p_2,p_3,p_4;\eps_1,\eps_3,\eps_4,\eps_5).
\end{eqnarray}
Similarly we can write for the prototype box amplitude,
\begin{eqnarray}
{\cal M}_B(p_1,p_2,p_3;\eps_1,\eps_2=p_2,\eps_3,\eps_{45}) &=& {\cal M}_T(p_1,p_2+p_3;\eps_1,\eps_3,\eps_{45}) 
  \nonumber \\
&& - {\cal M}_T(p_1+p_2,p_3;\eps_1,\eps_3,\eps_{45}). 
\end{eqnarray}
Here $\eps_{45}$ denotes the triple gluon vertex attached to the box/triangle diagram.
These Ward-Identities are shown diagrammatically in Fig. (\ref{fig:WI}).

 \newpage

\begin{figure}
\begin{minipage}[b]{0.5\linewidth}
\centering
\includegraphics[width=\textwidth]{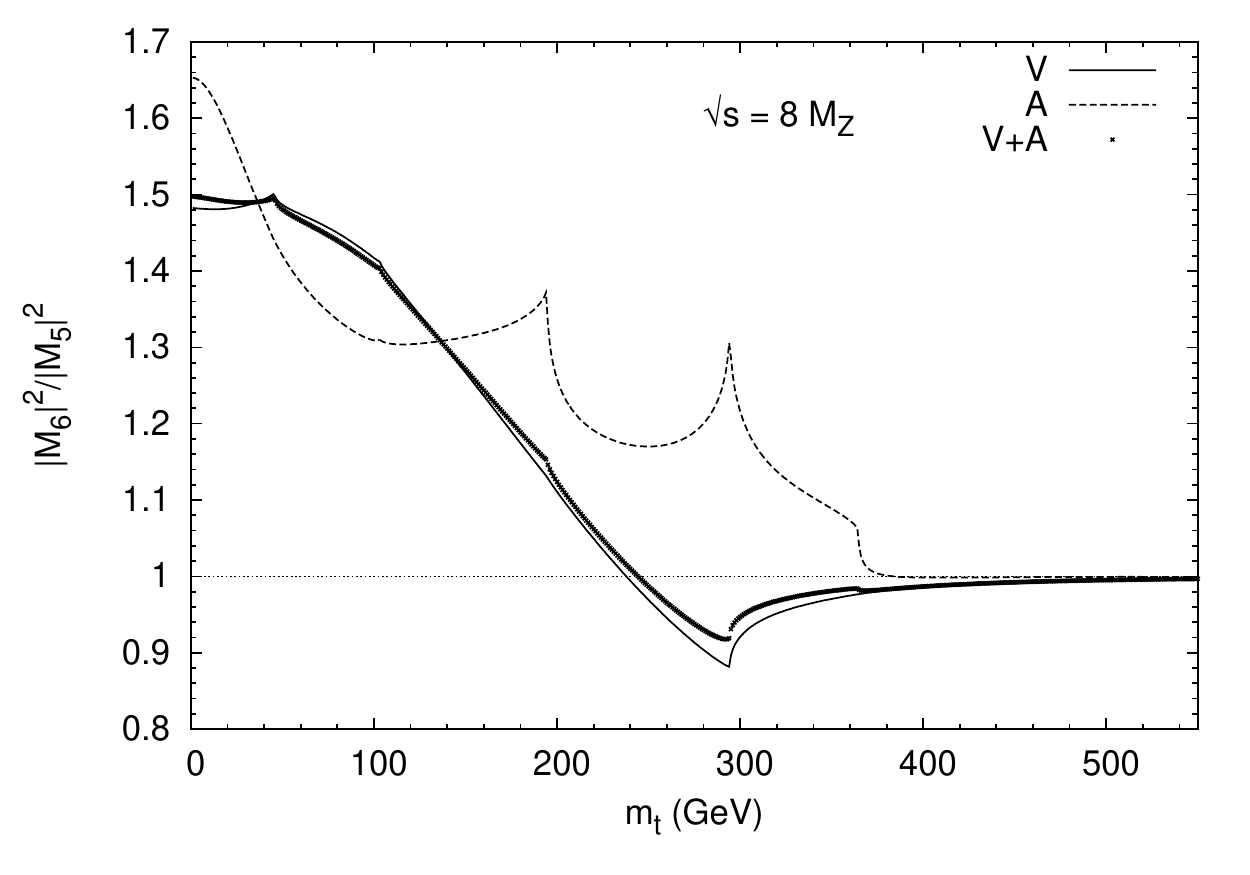}
 \caption{Decoupling of the top quark in the amplitude-squared for $\gpZ$. The vector and 
          the axial-vector contributions are shown separately.}
\label{fig:ampsq_mt}
\end{minipage}
\hspace{0.5cm}
\begin{minipage}[b]{0.5\linewidth}
\centering
\includegraphics[width=\textwidth]{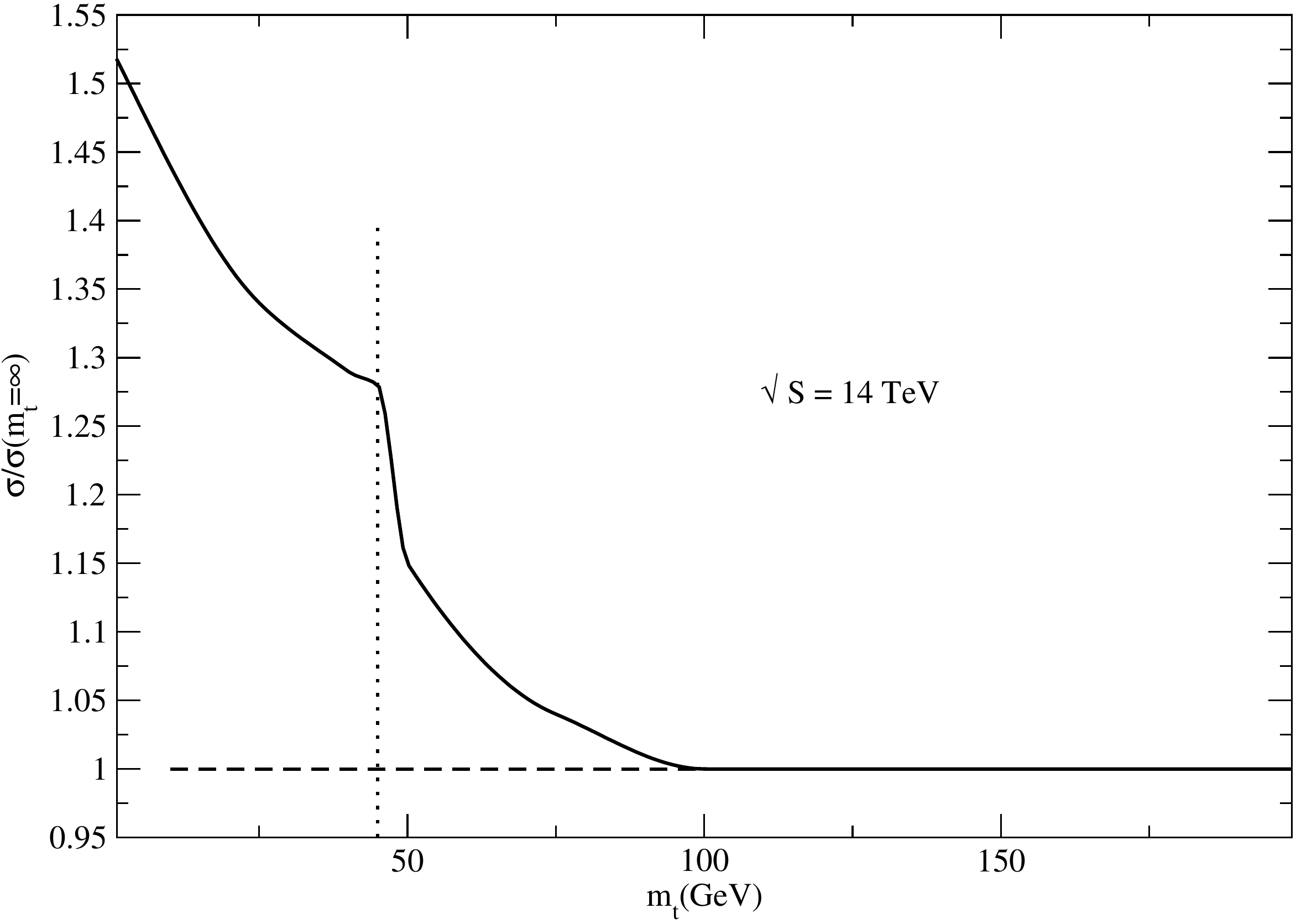}
 \caption{Decoupling of the top quark in the total cross section calculation for $\gpZ$.}
\label{fig:sigma_mt}
\end{minipage}
\end{figure}
\begin{figure}
\begin{minipage}[b]{0.5\linewidth}
\centering
\includegraphics[width=\textwidth]{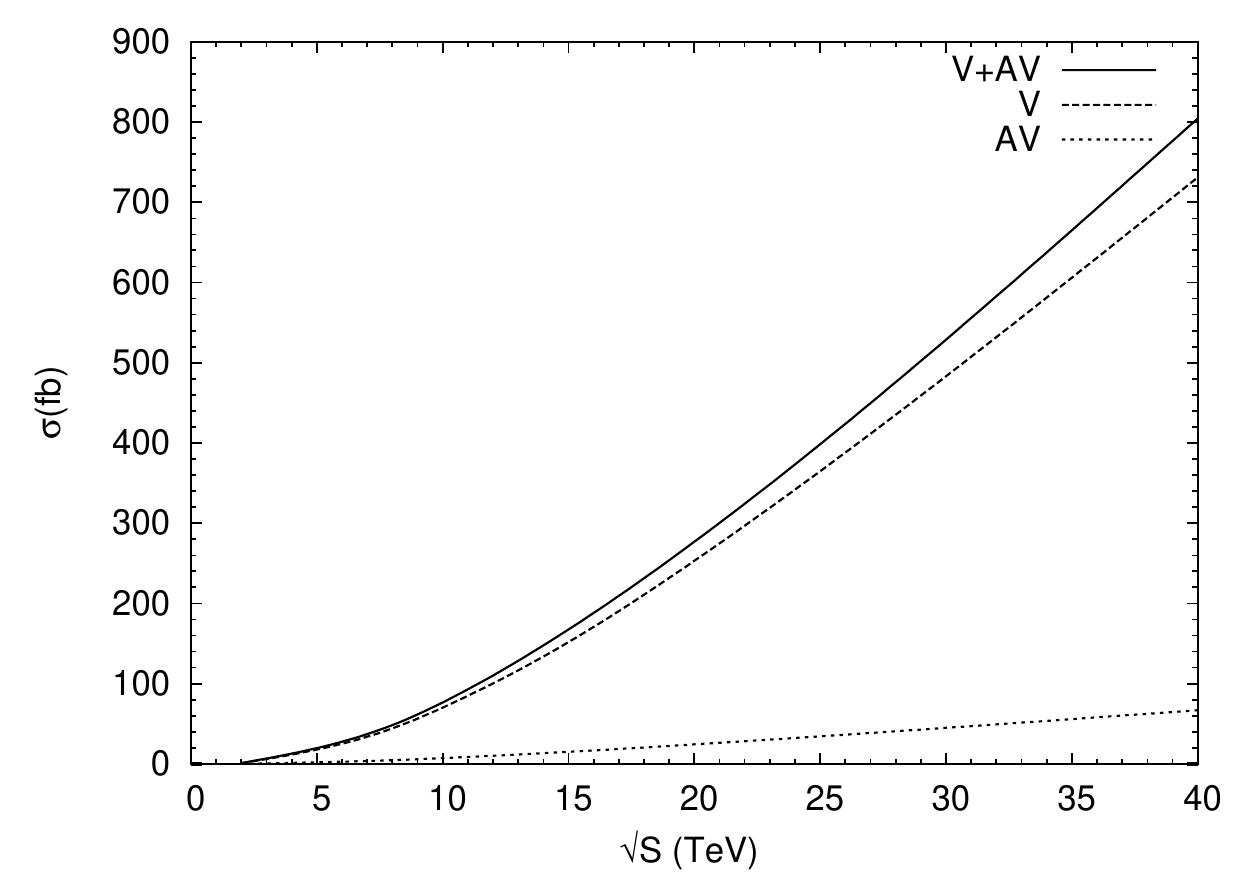}
 \caption{The vector and the axial-vector contributions towards the total cross section for $\gpZ$.}
\label{fig:sigma_cme-V+A}
\end{minipage}
\hspace{0.5cm}
\begin{minipage}[b]{0.5\linewidth}
\centering
\includegraphics[width=\textwidth]{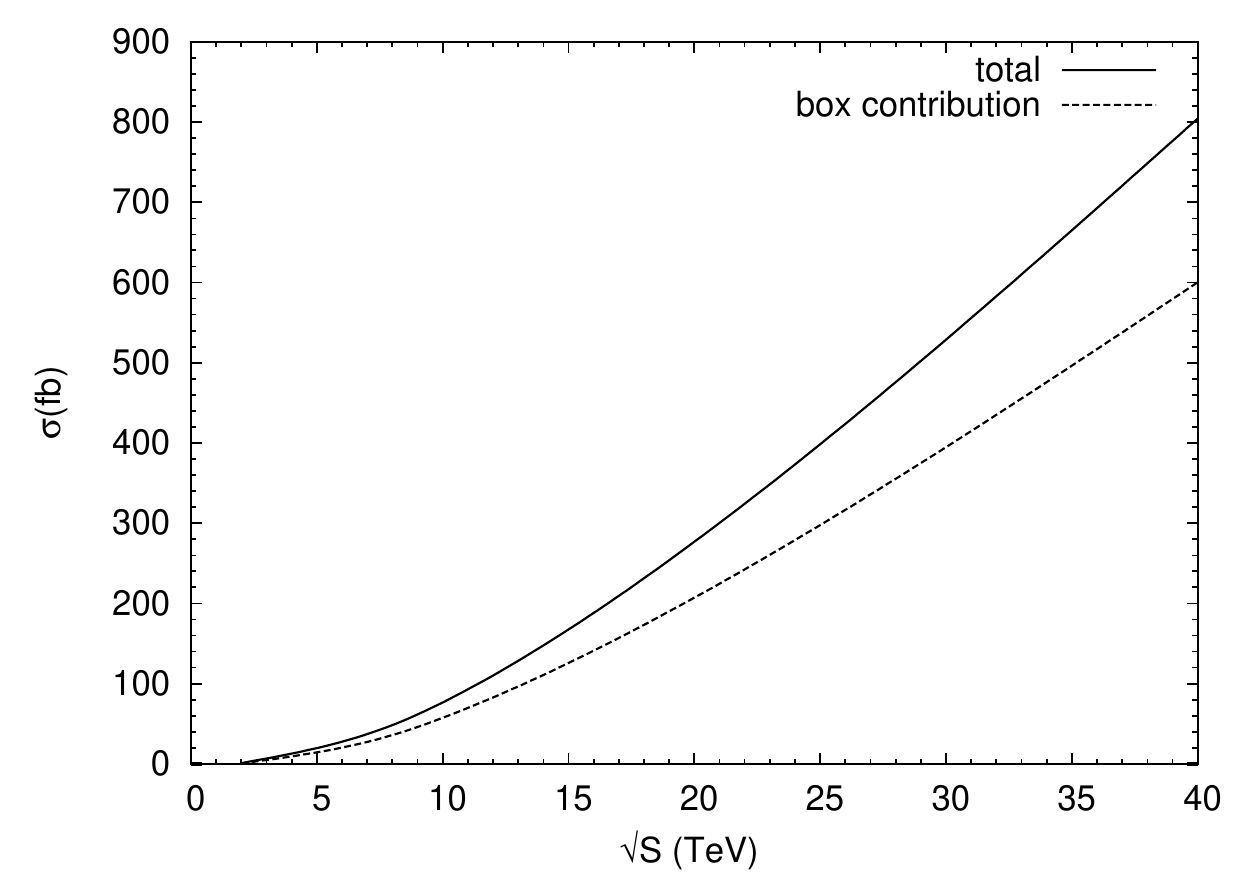}
 \caption{Contribution of the box amplitude towards the total cross section for $\gpZ$.}
\label{fig:sigma_cme-box}
\end{minipage}
\end{figure}
\begin{figure}[ht]
\begin{minipage}[b]{0.5\linewidth}
\centering
\includegraphics[width=\textwidth]{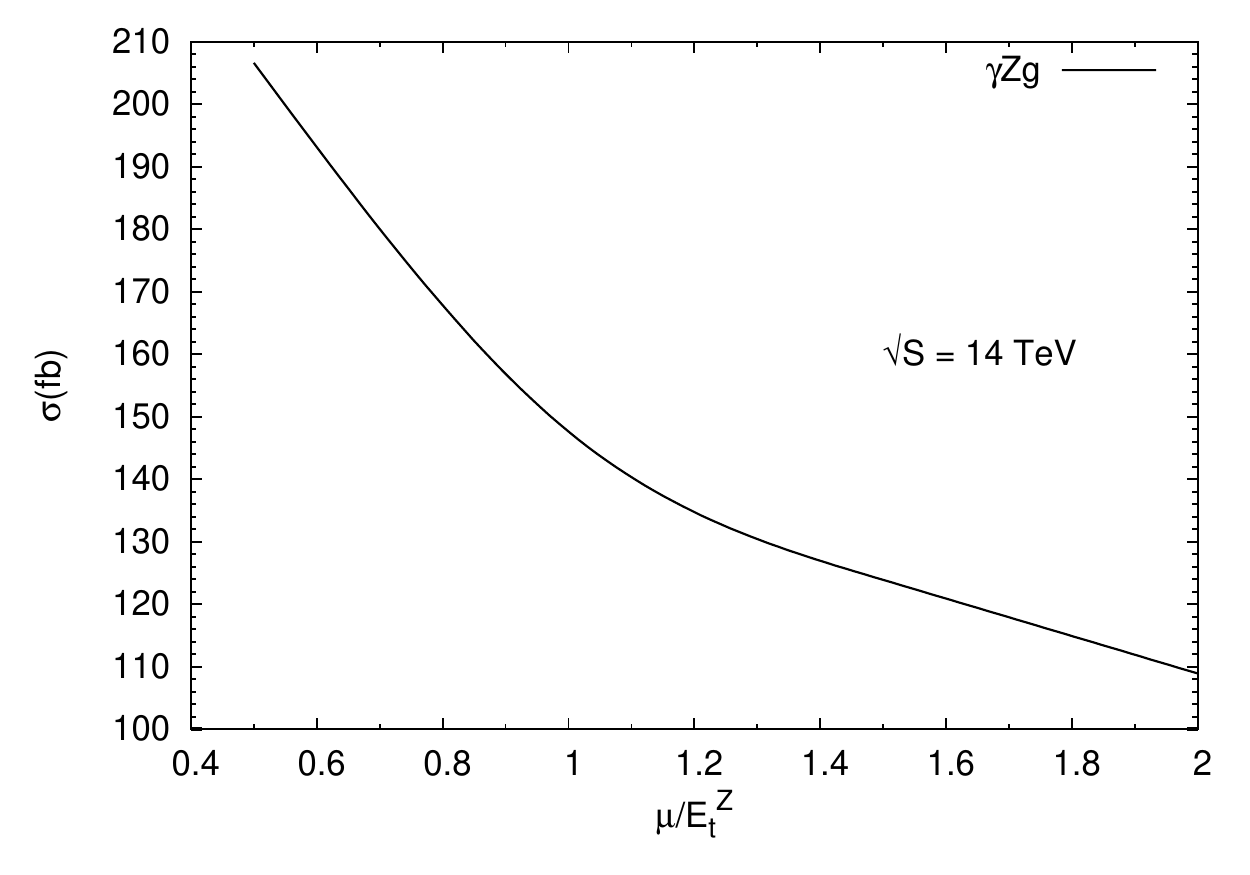}
 \caption{Variation of the total cross section for $\gpZ$ with the scale, $\mu = \mu_R = \mu_F$.}
\label{fig:sigma14Q}
\end{minipage}
\hspace{0.5cm}
\begin{minipage}[b]{0.5\linewidth}
\centering
\includegraphics[width=\textwidth]{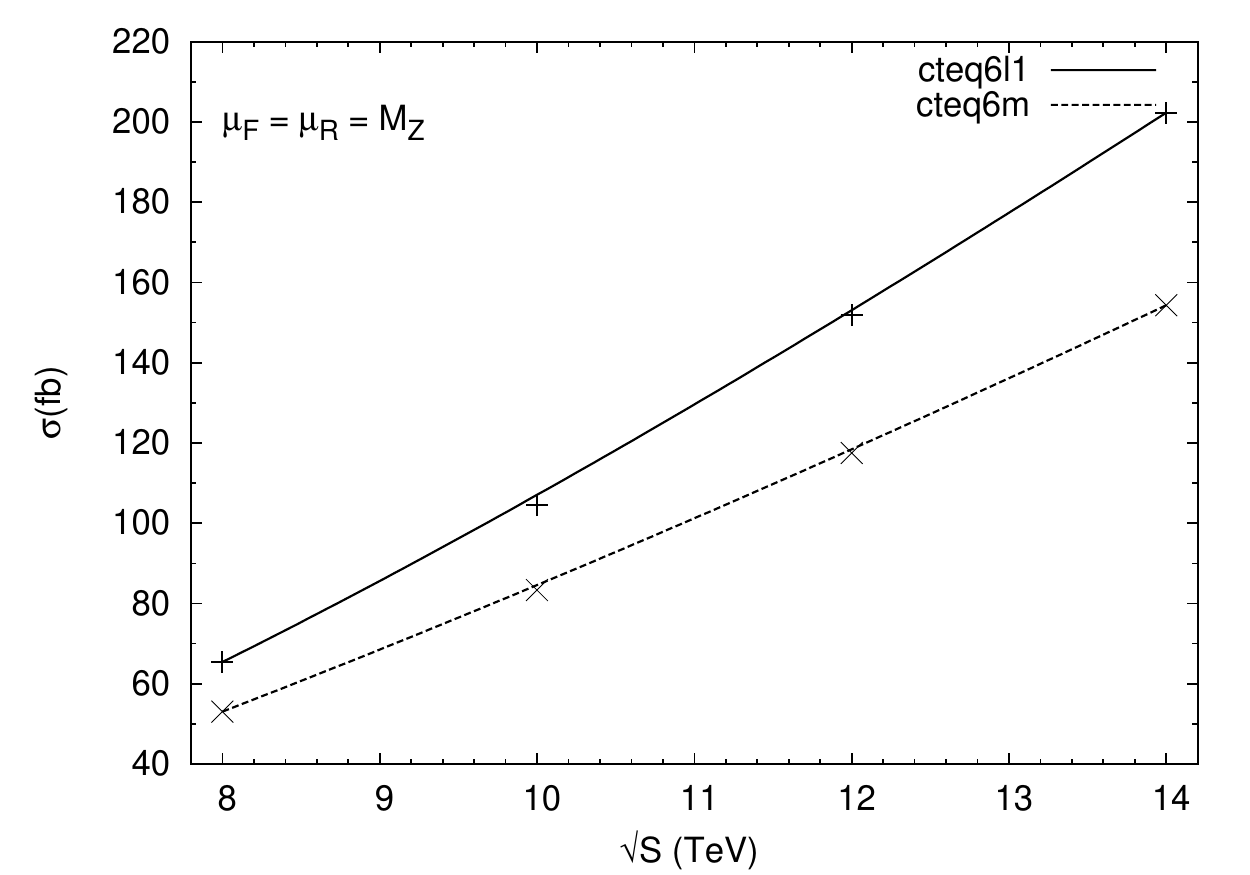}
 \caption{Dependence of the total cross section for $gg\to \gamma Z(\to l^+l^-) g$ on collider centre-of-mass energy.}
\label{fig:sigma_cme-gpz}
\end{minipage}
\end{figure}
\begin{figure}[ht]
\begin{minipage}[b]{0.5\linewidth}
\centering
\includegraphics[width=\textwidth]{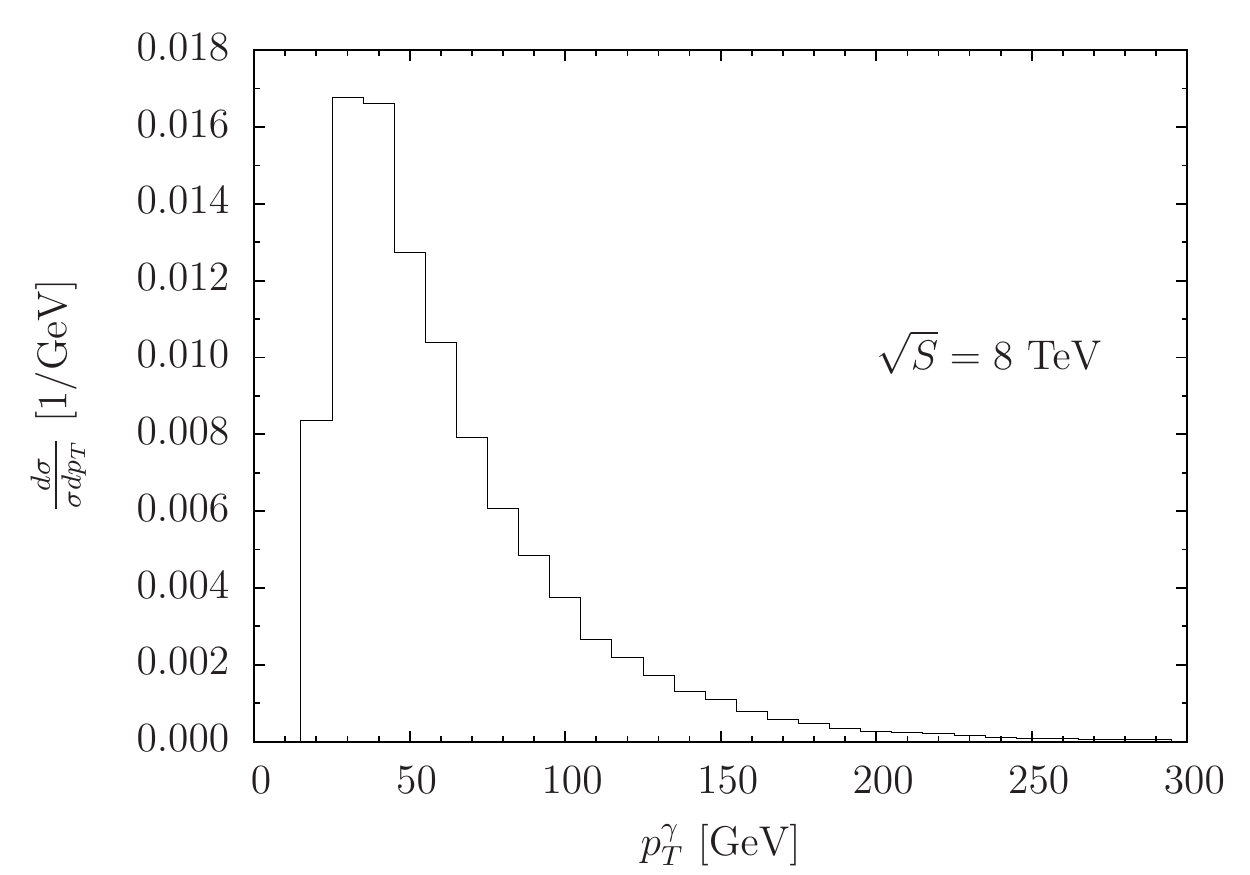}
 \caption{Transverse momentum distribution of $\gamma$ at 8 TeV centre-of-mass energy for $gg\to \gamma Z(\to l^+l^-) g$. }
\label{fig:ptp8}
\end{minipage}
\hspace{0.5cm}
\begin{minipage}[b]{0.5\linewidth}
\centering
\includegraphics[width=\textwidth]{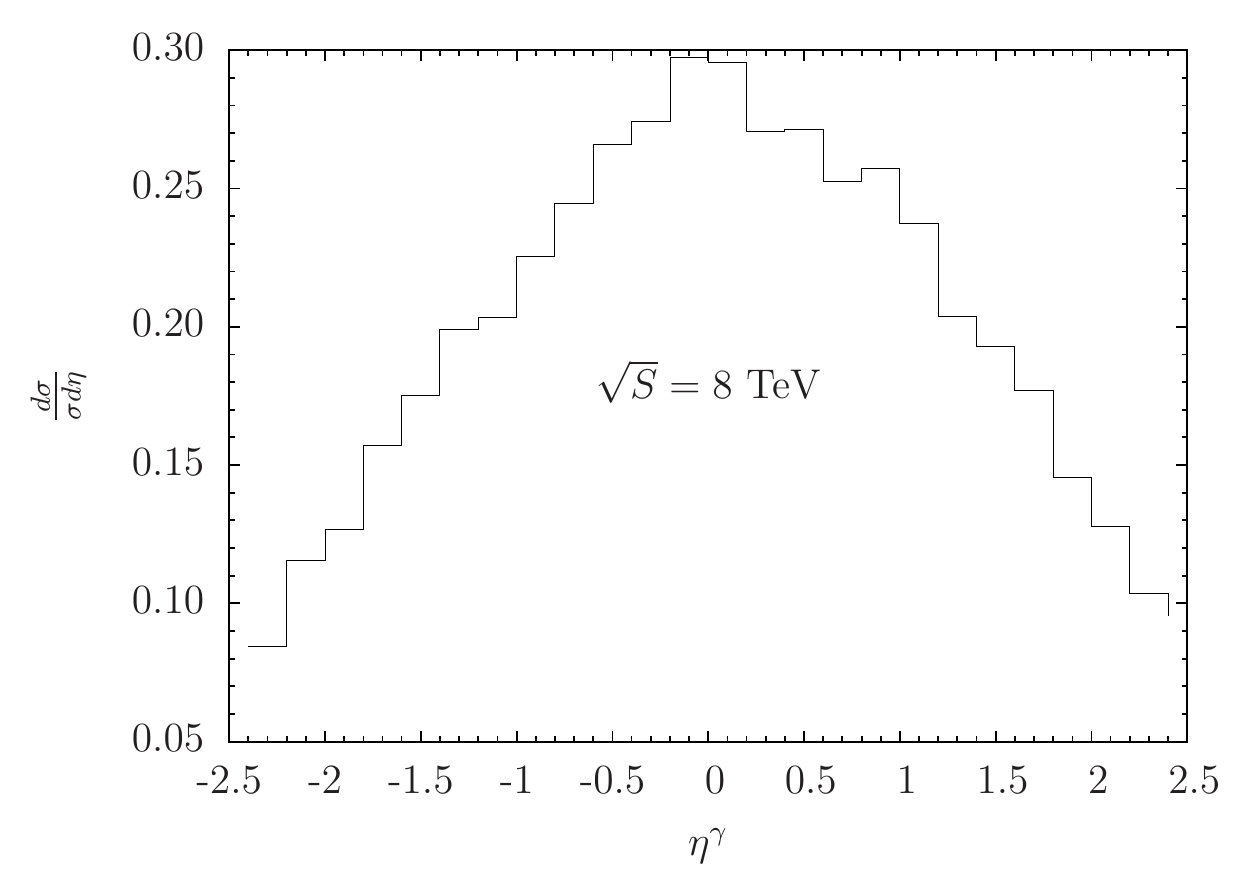}
 \caption{Rapidity distribution of $\gamma$ at 8 TeV centre-of-mass energy for $gg\to \gamma Z(\to l^+l^-) g$. }
\label{fig:etap8}
\end{minipage}
\end{figure}
\begin{figure}[ht]
\begin{minipage}[b]{0.5\linewidth}
\centering
\includegraphics[width=\textwidth]{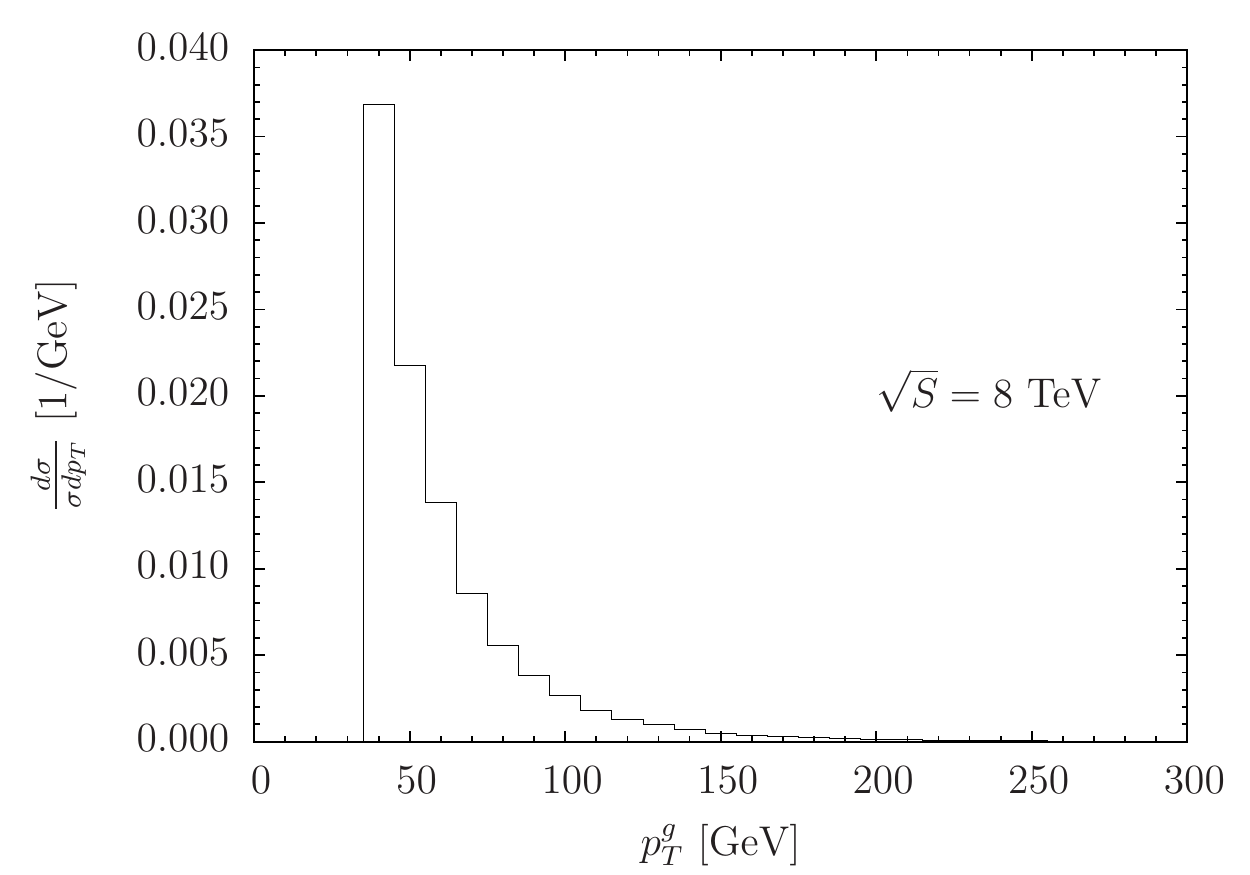}
 \caption{Transverse momentum distribution of the jet at 8 TeV centre-of-mass energy for $gg\to \gamma Z(\to l^+l^-) g$. }
\label{fig:ptg8}
\end{minipage}
\hspace{0.5cm}
\begin{minipage}[b]{0.5\linewidth}
\centering
\includegraphics[width=\textwidth]{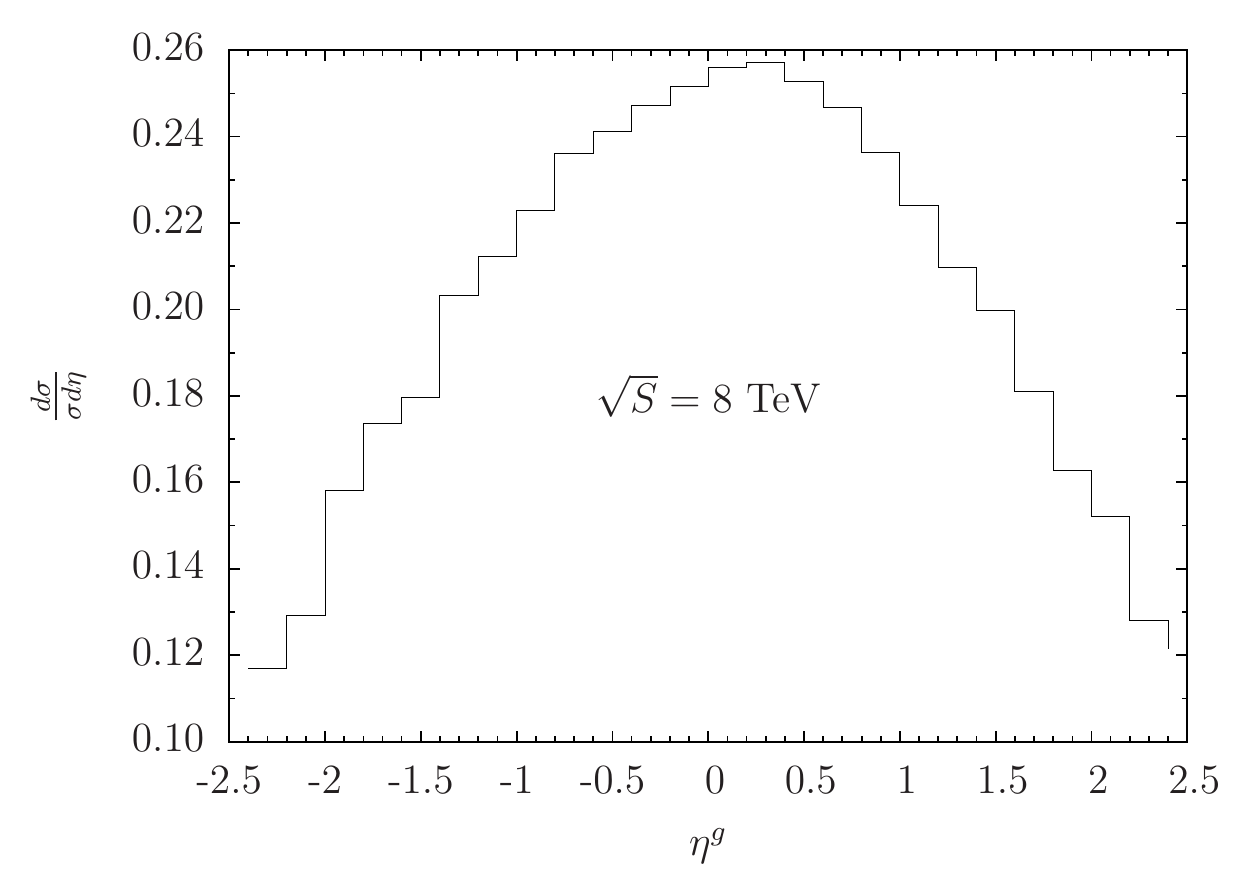}
 \caption{Rapidity distribution of the jet at 8 TeV centre-of-mass energy for $gg\to \gamma Z(\to l^+l^-) g$. }
\label{fig:etag8}
\end{minipage}
\end{figure}
\begin{figure}[ht]
\begin{minipage}[b]{0.5\linewidth}
\centering
 \includegraphics[width=\textwidth]{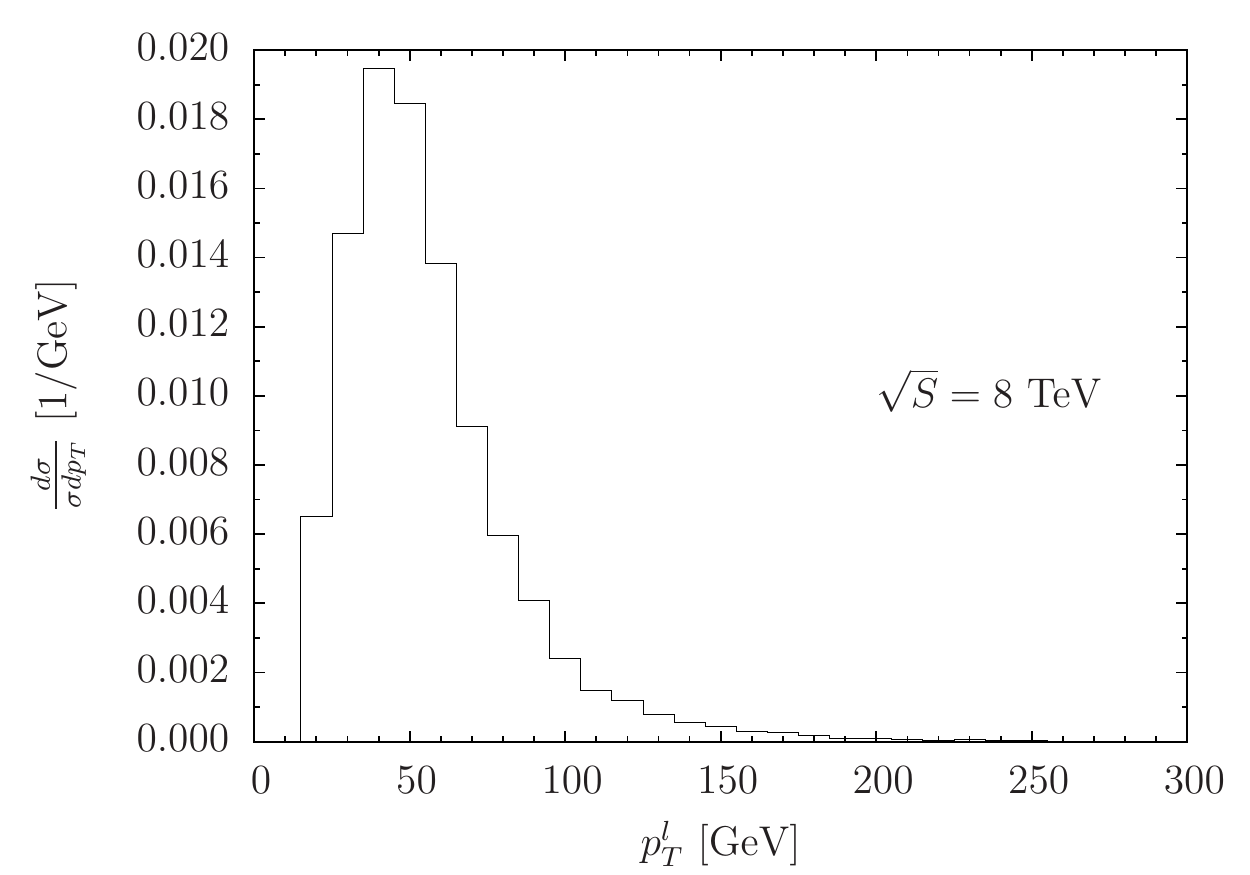}
 \caption{Transverse momentum distribution of lepton at 8 TeV centre-of-mass energy for $gg\to \gamma Z(\to l^+l^-) g$. }
\label{fig:ptl28}
\end{minipage}
\hspace{0.1cm}
\begin{minipage}[b]{0.5\linewidth}
\centering
\includegraphics[width=\textwidth]{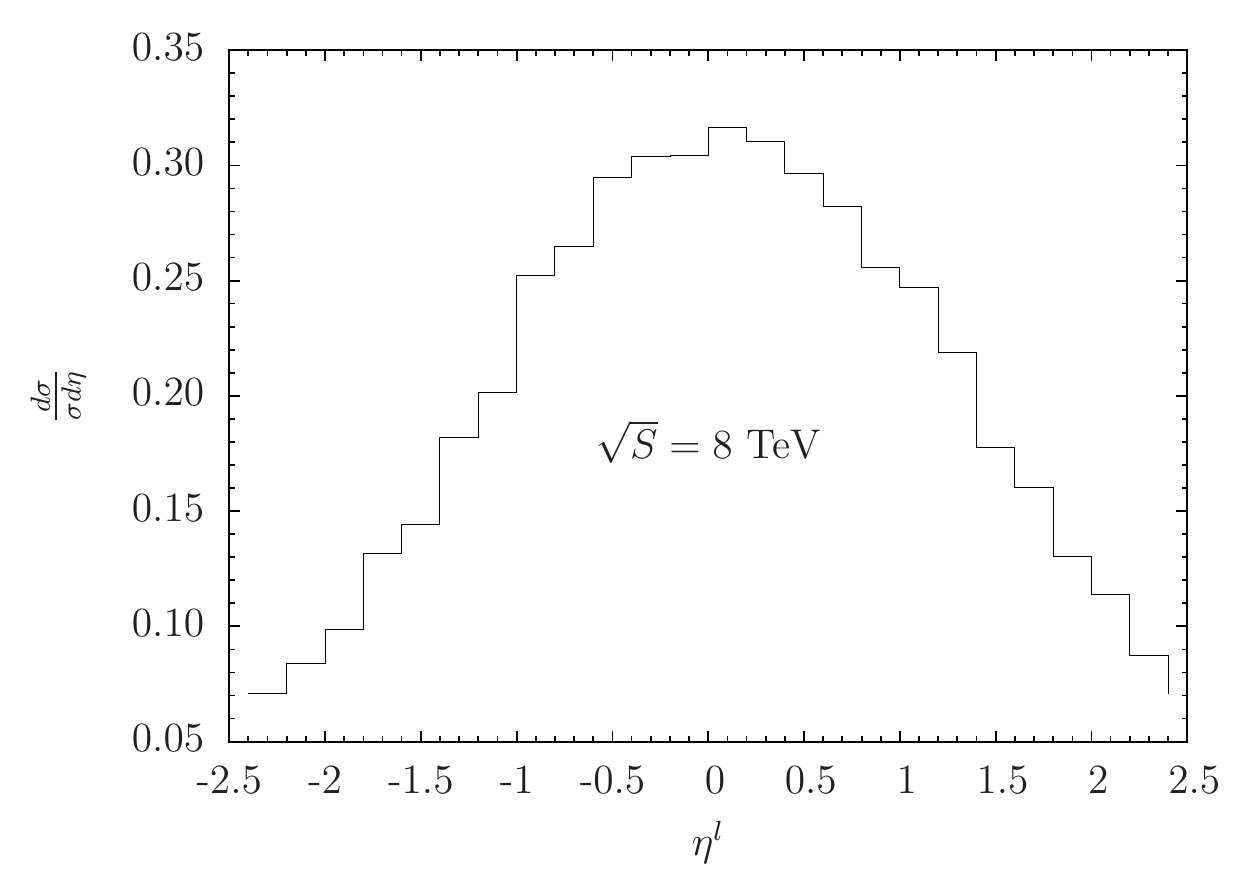}
 \caption{Rapidity distribution of lepton at 8 TeV centre-of-mass energy for $gg\to \gamma Z(\to l^+l^-) g$. }
\label{fig:etal28}
\end{minipage}
\end{figure}
\begin{figure}[ht]
\begin{minipage}[b]{0.5\linewidth}
\centering
\includegraphics[width=\textwidth]{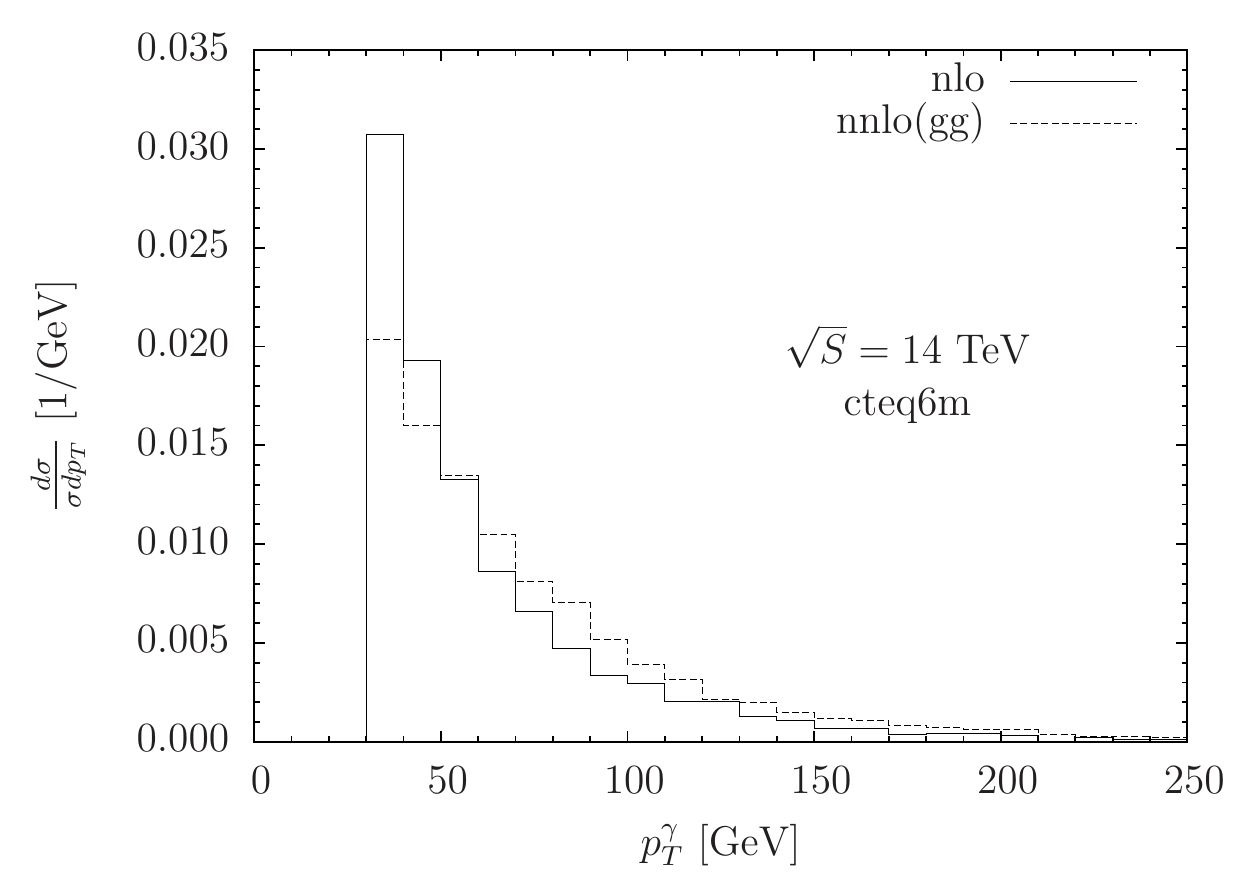}
 \caption{A comparison of normalized $p_T$ distributions of $\gamma$ at NLO and NNLO. The NLO
          distribution is obtained using MCFM.}
\label{fig:sigma14ptp30}
\end{minipage}
\hspace{0.5cm}
\begin{minipage}[b]{0.5\linewidth}
\centering
\includegraphics[width=\textwidth]{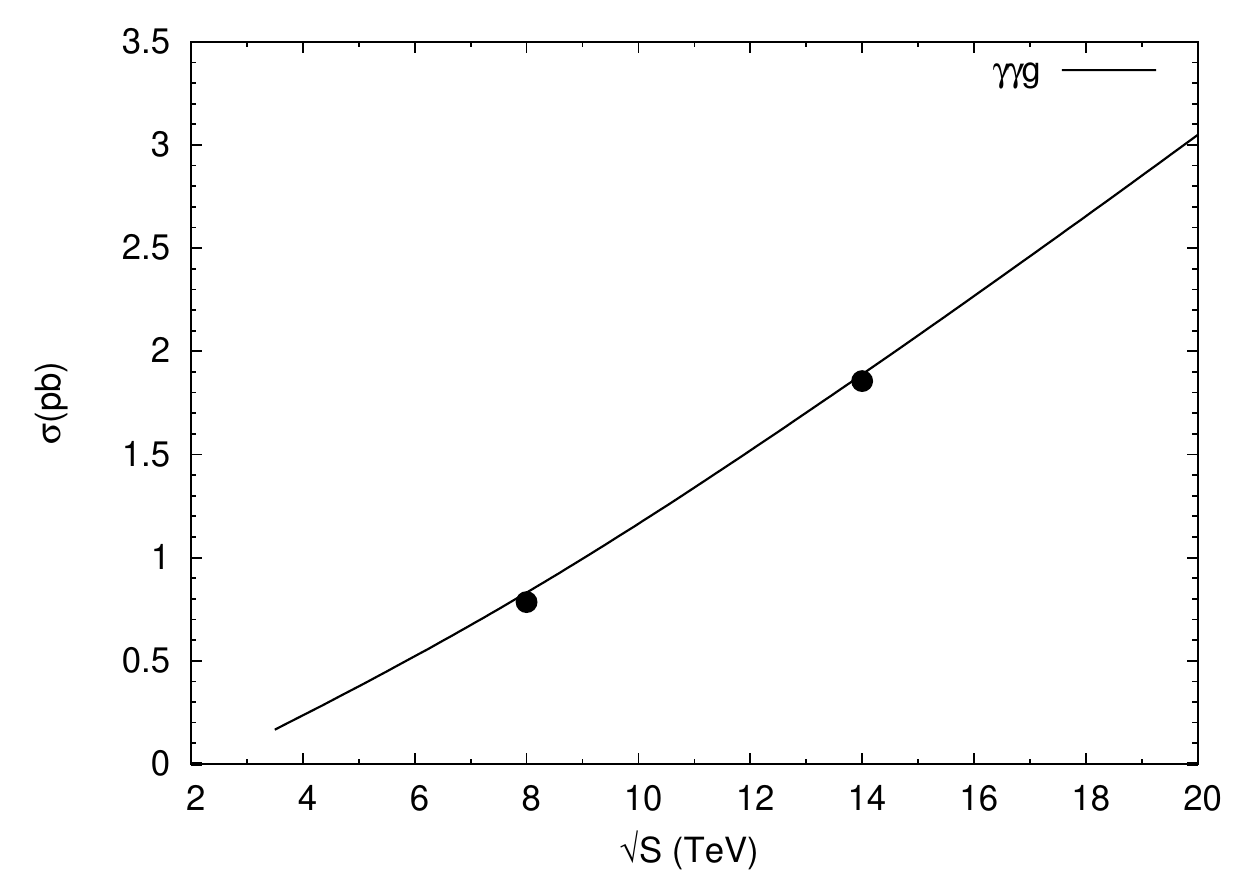}
 \caption{Variation of the total cross section with the collider centre-of-mass energy for $\ppg$. }
\label{fig:sigma_cme-ppg}
\end{minipage}
\end{figure}
\begin{figure}[ht]
\begin{minipage}[b]{0.5\linewidth}
\centering
\includegraphics[width=\textwidth]{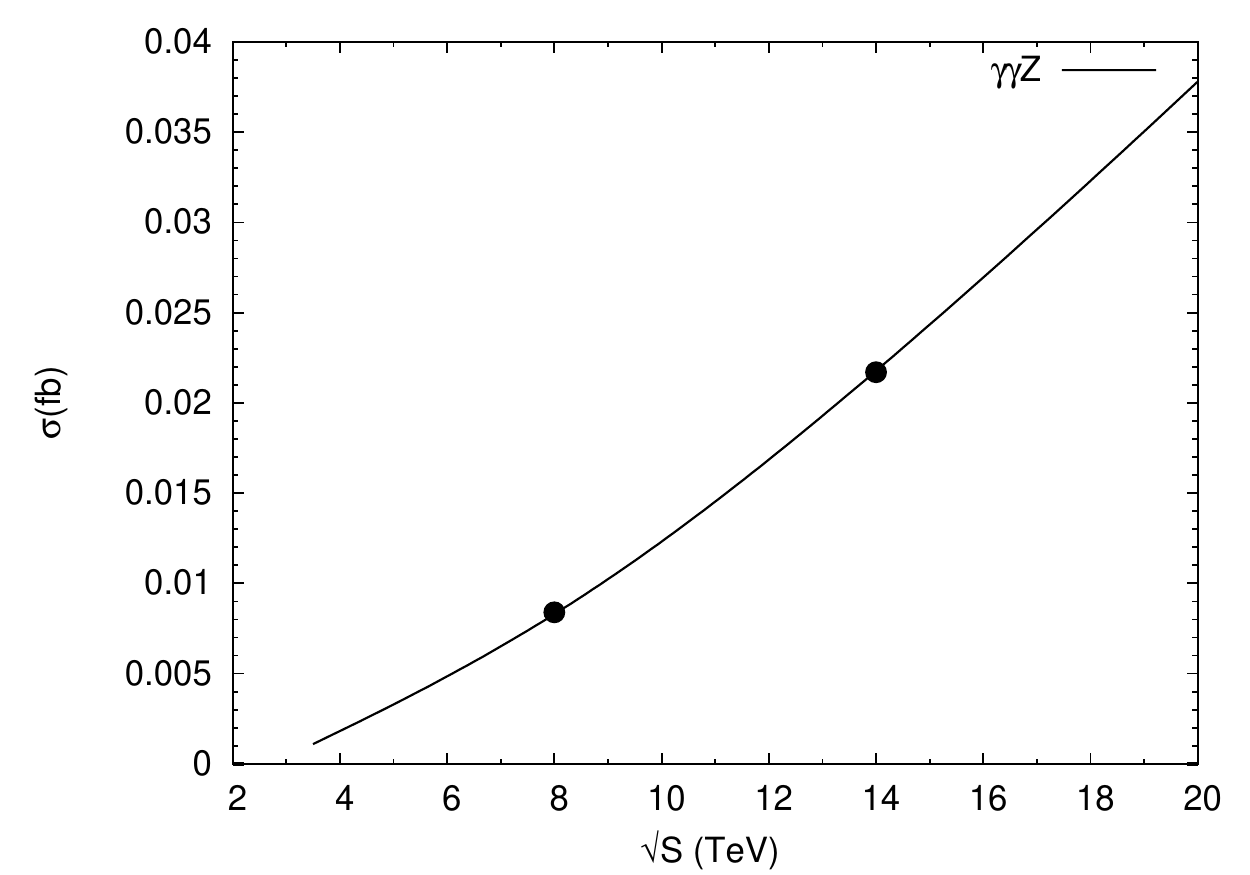}
 \caption{Variation of the total cross section with the collider centre-of-mass energy for $\ppZ$. }
\label{fig:sigma_cme-ppz}
\end{minipage}
\hspace{0.5cm}
\begin{minipage}[b]{0.5\linewidth}
\centering
\includegraphics[width=\textwidth]{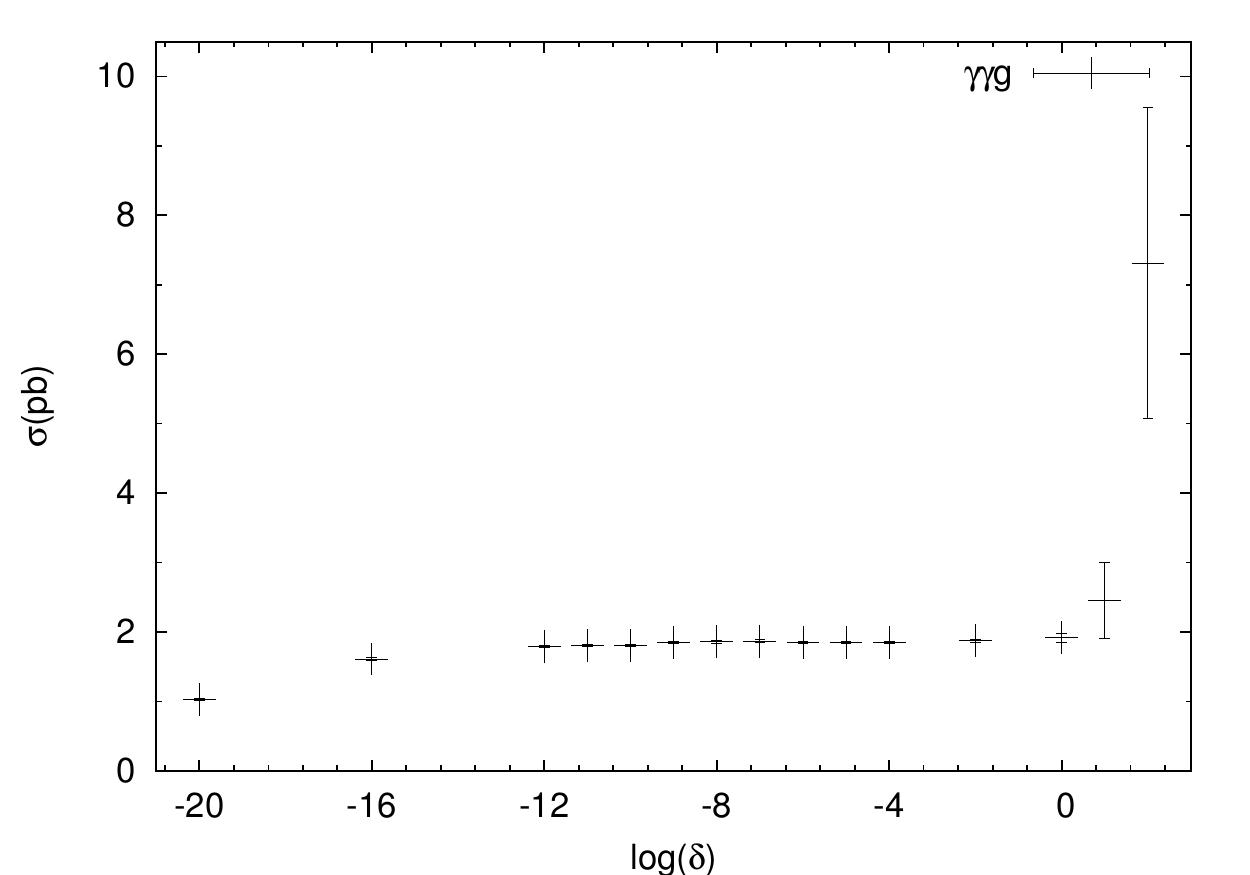}
\caption{Dependence of the total cross section on GI test cut parameter ($\delta$) for $\ppg$ at 14 TeV.}
\label{fig:sigma-cut}
\end{minipage}
\end{figure}
%
\begin{figure}[h!]
\begin{center}
\includegraphics [angle=0,width=0.8\linewidth] {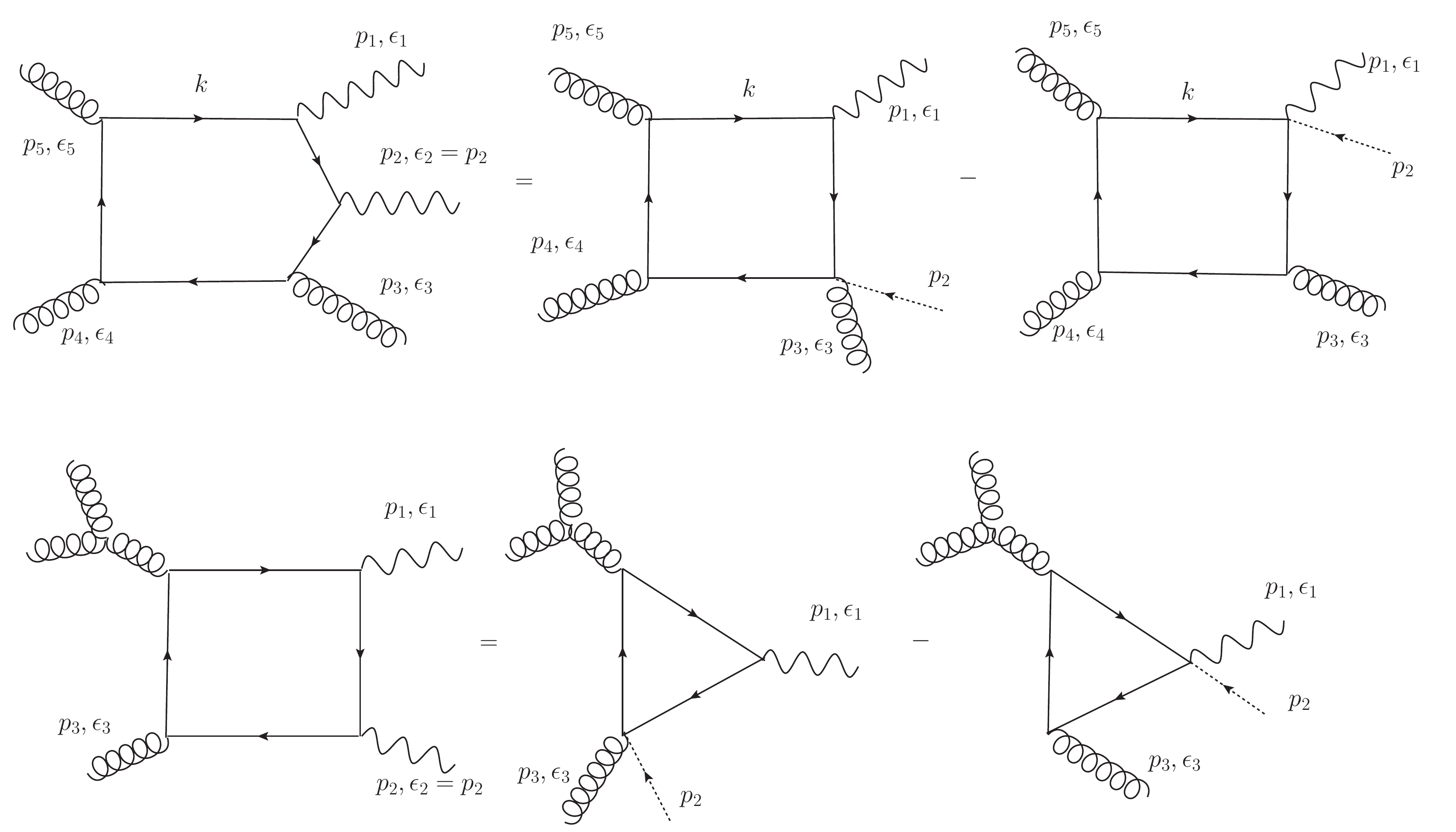}
\caption{Ward identities for pentagon and box diagrams. All the momenta are taken incoming. The 
 dotted lines take care of momentum insertion at relevant vertices.}
\label{fig:WI}
\end{center}
\end{figure}

\end{document}